\documentclass[default]{aastex701}

\newcommand\Ert{{\cal E}_{\rm t}}

\newcommand\rt{r_{\rm t}}

\newcommand\rz{r_{\rm n}}
\newcommand\st{s_{\rm t}}
\newcommand\xv{{\bf x}}
\newcommand\xvz{{\bf x}_0}
\newcommand\dtxv{d^3\xv}
\newcommand\diver{{\rm div}}

\begin{document}

\title{A (very) simple proof of the gravitational energy formula of polytropic spheres}

\author[orcid=0000-0002-5708-5274,sname='Luca Ciotti']{Luca Ciotti}
\affiliation{Dept. of Physics and Astronomy, via Piero Gobetti 93/2, University of Bologna - Italy}
\email[show]{luca.ciotti@unibo.it}  

\begin{abstract}
It is shown how the well-known formula for the gravitational energy of self-gravitating regular polytropes of finite mass can be obtained in an elementary way by using Gauss's divergence theorem and the Chandrasekhar virial tensor, without resorting to lengthy algebra, to specific properties of Lane-Emden functions, and to thermodynamics arguments, as is instead commonly found in standard treatises and in astrophysical literature. The present approach, due to its simplicity, can be particularly useful to students and researchers, and it can be easily applied to the study of more complicated polytropic structures.
\end{abstract}




\section{Introduction} 
\label{sec:intro}

The study of self-gravitating, polytropic gaseous spheres played a fundamental role in the development of the theory of stellar structure and evolution before the advent of computers (see \citealt{Chandra57}, hereafter C39), for example, explaining the White Dwarfs's (WDs) radius-mass relationship and leading to the discovery of the maximum mass limit for relativistically fully degenerate WDs. In Stellar Dynamics, self-gravitating collisionless stellar systems mathematically equivalent to gaseous polytropes are also studied (the so-called {\it stellar polytropes}, e.g., see \citealt{BT08,Ciotti21}, hereafter BT08 and C21), so that the following considerations apply to these systems as well. In these Notes the formula of the total gravitational energy of self-gravitating regular polytropes of finite mass and size is obtained in an extremely simple and transparent way, at variance with the lengthy and not-intuitive proofs usually found in the astrophysical literature.  

\section{Polytropes} 
\label{sec:material}

A quite complete account of the mathematical properties of polytropic spheres can be found in \cite{Horedt04}; here only the properties strictly necessary for the discussion are given. In a gaseous polytropic configuration the pressure and density are related as $p=K\rho^{\gamma}$, where $\gamma$ is the polytropic index, and $K=p_0/\rho_0^{\gamma}$ is a scaling constant, determined by the values of pressure and density at some position $\xvz$ in the system (e.g., the origin in the case of regular structures);
usually (but not necessarily) the equation of state of the perfect gas is assumed. If the system is in hydrostatic equilibrium in the gravitational potential $\phi$ (for the moment generic), then $\nabla p=\rho\nabla\Psi$, where $\Psi\equiv-\phi$ is the {\it relative potential}. The general solution for $\gamma>1$, obtained by line integration of the previous equation from $\xvz$ to a generic point $\xv$, and imposing non-negativity of the density, reads (e.g., see C21)
\begin{equation}
\rho=B\,\Phi^n\,\theta(\Phi),\quad 
n={1\over\gamma-1},\quad
\Phi\equiv\Psi-\Ert,
\label{eq:rhopoly}
\end{equation}
where $\theta$ is the Heaviside step function and $\Ert$ is the {\it truncation potential}, with
\begin{equation}
B = {1\over (n \gamma K )^n},\quad\Ert=\Psi_0 - {n \gamma p_0\over \rho_0}.
\label{eq:rhopolyn}
\end{equation}
Therefore, if $\Psi\to 0$ for $r\to\infty$\footnote{A potential vanishing at infinity does not necessarily imply a finite mass of the system producing it (e.g., see C21).}, and $\Ert >0$, then $\rho$ and $p$ vanish on the truncation surface defined by $\Phi=0$, where $\Psi=\Ert$; instead, for $\Ert\leq 0$ the system is spatially {\it untruncated}. 

Equation~(\ref{eq:rhopoly}) is fully general. In the self-gravitating case the potential is produced by the density distribution, so that $\rho$ and $\phi$ must be determined simultaneously. Restricting to spherical symmetry\footnote{From a fundamental result 
 \citep{Gidas79}, this restriction is not as arbitrary as it may appear.}, the Poisson equation to be solved for $\Phi$ reads 
\begin{equation}
{1\over r^2}{d\over dr}\left(r^2 {d\Phi\over dr}\right)=-4\pi G B \Phi^n\theta(\Phi),\quad \Phi (0)=\Phi_0,\quad
\Phi'(0)=0,
\label{eq:PoissonLE}
\end{equation}
where {\it regular} boundary conditions at the origin are adopted, as usual in applications\footnote{As Eq.~(\ref{eq:PoissonLE}) is non-linear, also {\it singular} solutions exist; pure power-law solutions 
\begin{equation}
\varphi=\left[{2(n-3)\over (n-1)^2}\right]^{1\over n-1}s^{2\over 1-n}
\end{equation}
can be easily found for $n>3$ (e.g., see C39).}. Equation~(\ref{eq:PoissonLE}) is cast in dimensionless form by introducing the scaled potential and radius as
\begin{equation}
 \varphi\equiv {\Phi\over\Phi_0},\quad
s\equiv {r\over\rz},\quad 
 \rz \equiv {1\over\sqrt{4\pi G B\Phi_0^{n-1}}},
\label{eq:PolyScaling}
\end{equation}
obtaining the well--known {\it Lane-Emden} equation (\citealt{Emden07,Fowler30}; C39):
\begin{equation}
{1\over s^2}{d\over ds}\left(s^2 {d\varphi\over ds}\right)=-
\varphi^n\theta(\varphi),\quad \varphi(0)=1,\quad \varphi'(0)=0;
\label{eq:LaneEmden}
\end{equation}
once the equation above is solved, all the physical properties of the resulting system are determined by fixing the values of $\Phi_0$ and $B$, so that 
\begin{equation}
\rho=B\Phi_0^n\varphi^n\theta(\varphi),\quad
p={B\Phi_0^{n\gamma}\varphi^{n\gamma}\over n\gamma}\theta(\varphi),\quad
{k_{\rm B}T\over\mu m_{\rm p}}={\Phi_0\varphi\over n\gamma}\theta(\varphi).
\label{eq:Polysol}
\end{equation}
where the last expression holds for a perfect gas.

For regular solutions with $n\geq 0$ it can be proven that (C39):
1) for $n<5$ the total mass $M$ of the density distribution is finite, and $\rho$ vanishes at a finite truncation radius $\rt=\rz\st$, with $\st$ implicitely defined by $\varphi(\st)=0$.
2) For $n=5$, $\rho$ is spatially untruncated but $M$ is still finite. 
3) For $n>5$, $\rho$ is spatially untruncated  (with $\varphi\to 0$ for $s\to\infty$), and $M$ is infinite. 
4) In addition to the elementary $n=0$ constant density case ($\st=\sqrt{6}$), only two (regular) analytical solutions exist, 
namely for $n=1$ (linear Helmholtz equation, with $\st=\pi$), and for $n=5$ (Schuster solution, with $\st=\infty$, also known in the astrophysical literature as the \citealt{Plummer11} model). It follows that for regular solutions, excluding the $n=0,1,5$ cases, $\varphi$ and $\st$ (for $n<5$) can only be obtained numerically.  In particular, from volume integration of Eq.~(\ref{eq:rhopoly}), 
\begin{equation}
  M=4\pi\rz^3\Phi_0^nB{\cal M}={\rz\Phi_0\over G}{\cal M},\quad
 {\cal M}\equiv\int_0^{\st}s^2\varphi^n\,ds=-\lim_{s\to\st}s^2{d\varphi\over
   ds};
 \label{eq:PolyMass}
\end{equation}
so that numerical determination of ${\cal M}$ is required in general (with the exception of ${\cal M}_0=2\sqrt{6}$, ${\cal M}_1=\pi$, and ${\cal M}_5=\sqrt{3}$). It should be now clear that in the regular self-gravitating case the truncation energy $\Ert$ appearing in Eq.~(\ref{eq:rhopoly}) is determined {\it after} the problem is solved: for $n\geq 5$ the system is untruncated and $\Ert$ can be set to zero, while for $n<5$
\begin{equation}
\Ert=\Psi(\rt)={GM\over\rt}=\Phi_0{{\cal M}\over\st},
\label{eq:polytrunc} 
\end{equation}
the relative gravitational potential at the density boundary surface. 

From previous arguments it is then quite remarkable that the gravitational energy of self-gravitating regular polytropes
can be written by the exact and beautifully simple formula 
\begin{equation}
U=-{3 M\Ert\over 5-n}=-{3 G M^2\over (5-n)\rt},\quad 0\leq n< 5,
\label{eq:polytW}
\end{equation}
first obtained by \cite{Betti80} and \cite{Ritter80} (see C39 for historical notes); of course, for $n\to 5$ the finite gravitational energy of the Plummer model is recovered, due to the balancing effect of $\rt\to\infty$ (see e.g. \citealt{Poveda58}).

\section{The proof}
\label{sec:proof}

As both $M$ and $\rt$ do not have an explicit expression, it should
not be a surprise that Eq.~(\ref{eq:polytW}) is usually proved with
lengthy algebraic manipulations based on the properties of the
Lane-Emden functions, and the use of thermodynamical identities, an
approach that somewhat obscures the origin of such a simple
expression. In fact, Eq.~(\ref{eq:polytW}) can be established
effortlessly from Gauss's divergence theorem and Chandrasekhar virial
tensor $W_{ij}$ with just a few passages, using Eq.~(\ref{eq:rhopoly})
{\it only}: curiously, such elementary approach does not seem to have
been mentioned (at least in the astronomical literature). The self-gravitational energy of density distribution $\rho(\xv)$ defined over some region $V$ (that can also be the whole space, here $V$ is the sphere of radius $\rt$), is given by the two equivalent expressions (\citealt{Chandra69}, see also BT08, C21)
\begin{equation}
W \equiv {\rm Tr}(W_{ij}) = \int_{V}\rho\langle\xv,\nabla\Psi\rangle\dtxv =  - {1\over 2}\int_V\rho\Psi\dtxv = U,
\label{eq:polyW1}
 \end{equation} 
where $\langle\cdot,\cdot\rangle$ indicates the standard inner product. A few 
elementary passages then follow from Eq.~(\ref{eq:rhopoly}):
\begin{eqnarray}
   U&=&
         B\int_V\Phi^n\langle\xv,\nabla\Phi\rangle\dtxv =
        {B\over n+1}\int_V\langle\xv,\nabla\Phi^{n+1}\rangle\dtxv ={B\over n+1}\int_V\left[\diver (\xv\Phi^{n+1})-3\Phi^{n+1}\right]\dtxv\cr\cr
   &=& -{3\over n+1}\int_V\rho\Phi\dtxv = {6U\over n+1}+{3M\Ert\over n+1}.
       \label{eq:polyW2}
 \end{eqnarray}
The last integral in the first line is obtained from the identity $\diver (\xv f) = 3 f + \langle\xv,\nabla f\rangle$ (an integration by parts), so that from the divergence theorem with $\Phi(\rt)=0$ the first contribution vanishes, while in the second term $B\Phi^{n+1}=\rho\Phi$ over the region $V$. The last integral is finally evaluated from $\Phi=\Psi - \Ert$, considering the last identity in Eq.~(\ref{eq:polyW1}), and that $\int_V\rho\dtxv=M$: solving Eq.~(\ref{eq:polyW2}) for $U$ concludes the proof.
 
\section{Conclusions}
\label{sec:concl}

It is shown how the formula expressing the gravitational energy of (regular) self-gravitating polytropic spheres with $n<5$ can be obtained in a few elementary passages just using 1) the functional form of density expressed in terms of the potential, 2) Gauss's divergence theorem, and 3) Chandrasekhar's virial tensor. At variance with the common derivation that can be found in the standard astronomical references on polytropic configurations, the proof does not use properties of the Lane-Emden functions and thermodynamical identities, nor does it require the memorization of several non-obvious integral manipulations, so that it can be useful to students and in general in research work. For example, the present approach shows immediately that in a polytropic distribution $\rho$ of total mass $M$ and truncation potential $\Ert$, at equilibrium in its own potential $\phi_{\rm self}$ plus an ``externally'' imposed potential $\phi_{\rm ext}$ (not necessarily spherically symmetric, so that also $\rho$ and $\phi_{\rm self}$ will not be spherically symmetric), the identity
\begin{equation}
{n+1\over 3}W_{\rm tot} = 2 U_{\rm self} + U_{\rm ext} + M\Ert,\quad {\rm i.e.}\quad 
{5-n\over 3}U_{\rm self} = - M\Ert +{n+1\over 3}W_{\rm ext} - U_{\rm ext}
\end{equation}
holds, where $U_{\rm ext}=\int_V\rho\phi_{\rm ext}\dtxv$, and $W_{\rm tot}=U_{\rm self} + W_{\rm ext}$ (e.g., see C21).

\bibliography{polyw}{}

\begin{thebibliography}{}
\expandafter\ifx\csname natexlab\endcsname\relax\def\natexlab#1{#1}\fi
\providecommand{\url}[1]{\href{#1}{#1}}
\providecommand{\dodoi}[1]{doi:~\href{http://doi.org/#1}{\nolinkurl{#1}}}
\providecommand{\doeprint}[1]{\href{http://ascl.net/#1}{\nolinkurl{http://ascl.net/#1}}}
\providecommand{\doarXiv}[1]{\href{https://arxiv.org/abs/#1}{\nolinkurl{https://arxiv.org/abs/#1}}}

\bibitem[{E. {Betti}(1880){Betti}}]{Betti80}
{Betti}, E. 1880, \bibinfo{title}{{Sopra l'Equilibrio di una Massa di Gaz
  Perfetto Isolata nello Spazio},} Nuovo Cimento, 7, 26

\bibitem[{J. {Binney} \& S. {Tremaine}(2008){Binney} \& {Tremaine}}]{BT08}
{Binney}, J., \& {Tremaine}, S. 2008, {Galactic Dynamics (BT08)}, 2nd edn.
  (Princeton University Press)

\bibitem[{S. {Chandrasekhar}(1939){Chandrasekhar}}]{Chandra57}
{Chandrasekhar}, S. 1939, {An Introduction to the Study of Stellar Structure
  (C39)} (Dover Publications)

\bibitem[{S. {Chandrasekhar}(1969){Chandrasekhar}}]{Chandra69}
{Chandrasekhar}, S. 1969, {Ellipsoidal Figures of Equilibrium} (Dover
  Publications)

\bibitem[{L. {Ciotti}(2021){Ciotti}}]{Ciotti21}
{Ciotti}, L. 2021, {Introduction to Stellar Dynamics (C21)} (Cambridge
  University Press)

\bibitem[{R. {Emden}(1907){Emden}}]{Emden07}
{Emden}, R. 1907, {Gaskugeln} (Teubner)

\bibitem[{R. {Fowler}(1930){Fowler}}]{Fowler30}
{Fowler}, R. 1930, \bibinfo{title}{{The Solutions of Emden's and Similar
  Differential Equations},} MNRAS, 91, 63

\bibitem[{B. {Gidas} {et~al.}(1979){Gidas}, {Ni}, \& {Nirenberg}}]{Gidas79}
{Gidas}, B., {Ni}, W.-M., \& {Nirenberg}, L. 1979, \bibinfo{title}{{Symmetry
  and Related Properties via the Maximum Principle},} Commun. Math. Phys., 68,
  209

\bibitem[{G. {Horedt}(2004){Horedt}}]{Horedt04}
{Horedt}, G. 2004, {Polytropes. Applications in Astrophysics and Related
  Fields}, Vol. 306 (Springer)

\bibitem[{H. {Plummer}(1911){Plummer}}]{Plummer11}
{Plummer}, H. 1911, \bibinfo{title}{{On the Problem of Distribution in Globular
  Star Clusters},} MNRAS, 71, 460

\bibitem[{A. {Poveda}(1958){Poveda}}]{Poveda58}
{Poveda}, A. 1958, \bibinfo{title}{{La energía potencial de la esfera
  politrópica n=5},} Boletín de los Observatorios de Tonantzintla y Tacubaya,
  17, 8

\bibitem[{A. {Ritter}(1880){Ritter}}]{Ritter80}
{Ritter}, A. 1880, \bibinfo{title}{{Untersuchungen uber die Hohe der Atmosphare
  und die Constitution gasformiger Weltkorper},} Wiedemann Annalen, 11, 332

\end{thebibliography}
\bibliographystyle{aasjournalv7}


\end{document}